\documentclass[journal]{IEEEtran}
\usepackage{cite}
\usepackage[pdftex]{graphicx}
\usepackage{amsmath}
\usepackage{amssymb}
\usepackage{textcomp}
\usepackage{amsfonts}
\usepackage{algpseudocode}
\usepackage{algorithm}
\usepackage{array}
\usepackage[caption=false,font=footnotesize]{subfig}

\usepackage{multirow}
\usepackage{makecell}
\usepackage{stfloats}
\usepackage{url}

\usepackage{cite}
\usepackage{balance}
\newcommand{\minitab}[2][l]{\begin{tabular}{#1}#2\end{tabular}}

\begin{document} 
%
\title{ACGAN-based Data Augmentation Integrated with Long-term Scalogram for Acoustic Scene Classification}
%
%
%

\author{Hangting~Chen,
	Zuozhen~Liu,
	Zongming~Liu,
	and~Pengyuan~Zhang,~\IEEEmembership{Member,~IEEE}
	\thanks{Manuscript received September 25, 2019. This work is partially supported by the National Key Research and Development Program (Nos. 2018YFC0823401, 2018YFC0823402, 2018YFC0823405, 2018YFC0823400), National Natural Science Foundation of China (Nos. 11590772, 11590774, 11590770), the Key Science and Technology Project of the Xinjiang Uygur Autonomous Region (No.2016A03007-1).(Corresponding author: Pengyuan Zhang.)}
	\thanks{Hangting Chen, Zuozhen Liu, Zongming Liu, Pengyuan Zhang are with the Institute of Acoustics, Chinese Academy of Sciences, Beijing 100190, China, and also with the University of Chinese Academy of Sciences, Beijing 100049, China (email: chenhangting@hccl.ioa.ac.cn, zhangpengyuan@hccl.ioa.ac.cn).}
}
%
%

\markboth{Journal of \LaTeX\ Class Files,~Vol.~14, No.~8, August~2015}%
{Shell \MakeLowercase{\textit{et al.}}: Bare Demo of IEEEtran.cls for IEEE Journals}
%



\maketitle

\begin{abstract}
In acoustic scene classification (ASC), acoustic features play a crucial role in the extraction of scene information, which can be stored over different time scales. Moreover, the limited size of the dataset may lead to a biased model with a poor performance for records from unseen cities and confusing scene classes. In order to overcome this, we propose a long-term wavelet feature that requires a lower storage capacity and can be classified faster and more accurately compared with classic Mel filter bank coefficients (FBank). This feature can be extracted with predefined wavelet scales similar to the FBank. Furthermore, a novel data augmentation scheme based on generative adversarial neural networks with auxiliary classifiers (ACGANs) is adopted to improve the generalization of the ASC systems. The scheme, which contains ACGANs and a sample filter, extends the database iteratively by splitting the dataset, training the ACGANs and subsequently filtering samples. Experiments were conducted on datasets from the Detection and Classification of Acoustic Scenes and Events (DCASE) challenges. The results on the DCASE19 dataset demonstrate the improved performance of the proposed techniques compared with the classic FBank classifier. Moreover, the proposed fusion system achieved first place in the DCASE19 competition and surpassed the top accuracies on the DCASE17 dataset.
\end{abstract}

\begin{IEEEkeywords}
DCASE, acoustic scene classification, scalogram, generative adversarial network.
\end{IEEEkeywords}

%
\IEEEpeerreviewmaketitle

\section{Introduction}

\IEEEPARstart{E}{nvironmental} sound carries a large amount of information related to the surroundings. Humans are able to understand the general context of the acoustic scene by listening to sounds and can automatically adapt the auditory systems. However, enabling devices to sense the environment is a challenging task of machine listening and computational auditory scene analysis \cite{BarchiesiAcoustic}. Acoustic scene classification (ASC) is one of the key topics required to improve the understanding of sound. It aims to classify sound into one of several predefined classes, e.g., park, office, library \cite{Mesaros2016TUT}\cite{Mesaros2017}. Obtaining key information of the environment has huge potential in audio-context-based searching \cite{bugalho2009detecting} and the manufacturing of context-aware devices \cite{eronen2005audio}. Detection and Classification of Acoustic Scenes and Events (DCASE) challenges are organized annually by the IEEE Audio and Acoustic Signal Processing (AASP) Technical Committee. Furthermore, the DCASE challenges have released some of the largest datasets in the field.

Real-world acoustic scene information is generally stored in background sounds, with variations in the time scale for cues. For example, the pitch and timbre are observed at a scale of milliseconds, while rhythm is of seconds \cite{Joakim2013Deep}. Figure 1 plots the short-time Fourier transform (STFT) spectra of audios recorded at an airport and a metro. In particular, the sound from the airport contains short-term footsteps located sparsely in the frequency axis. The sound from the metro exhibits the rhythm of the carriage advancing, which covers a time scale of hundreds of milliseconds and fills all the high and a part of the low frequencies. However, most of the conventional features, such as the Mel-scale filter bank coefficients (FBank) \cite{Stevens1937A}, are short-term and designed for tasks such as speech recognition. More specifically, they describe signals with many fluctuations, suitable for short-term phoneme recognition \cite{davis1980comparison} but inappropriate for depicting acoustic scenes. We assume that for ASC tasks, the scene cues are distributed along time at multiple scales, thus long-term features can act as a more accurate representation of background information. Long-term features still have the ability to describe short-term information by averaging over a long-term window, as long as the short-term cues are relatively distributed in a consistent manner within the spectra. Wavelet transform is able to model the signal at multiple resolutions \cite{mallat1999wavelet}. In our previous work, we demonstrated that by adopting a long-term window for the wavelet filters, the convolutional wavelet transform was able to outperform the FBank features \cite{chen2018deep}. The wavelet filters were applied directly to the STFT spectra, achieving a more accurate prediction compared to short-term features \cite{chen2019audio}. In this paper, we uncover the design of the wavelet filters and compare different settings to give a detailed analysis under various classifiers.

\begin{figure}[tb]
	\centering
	\includegraphics[width=\linewidth]{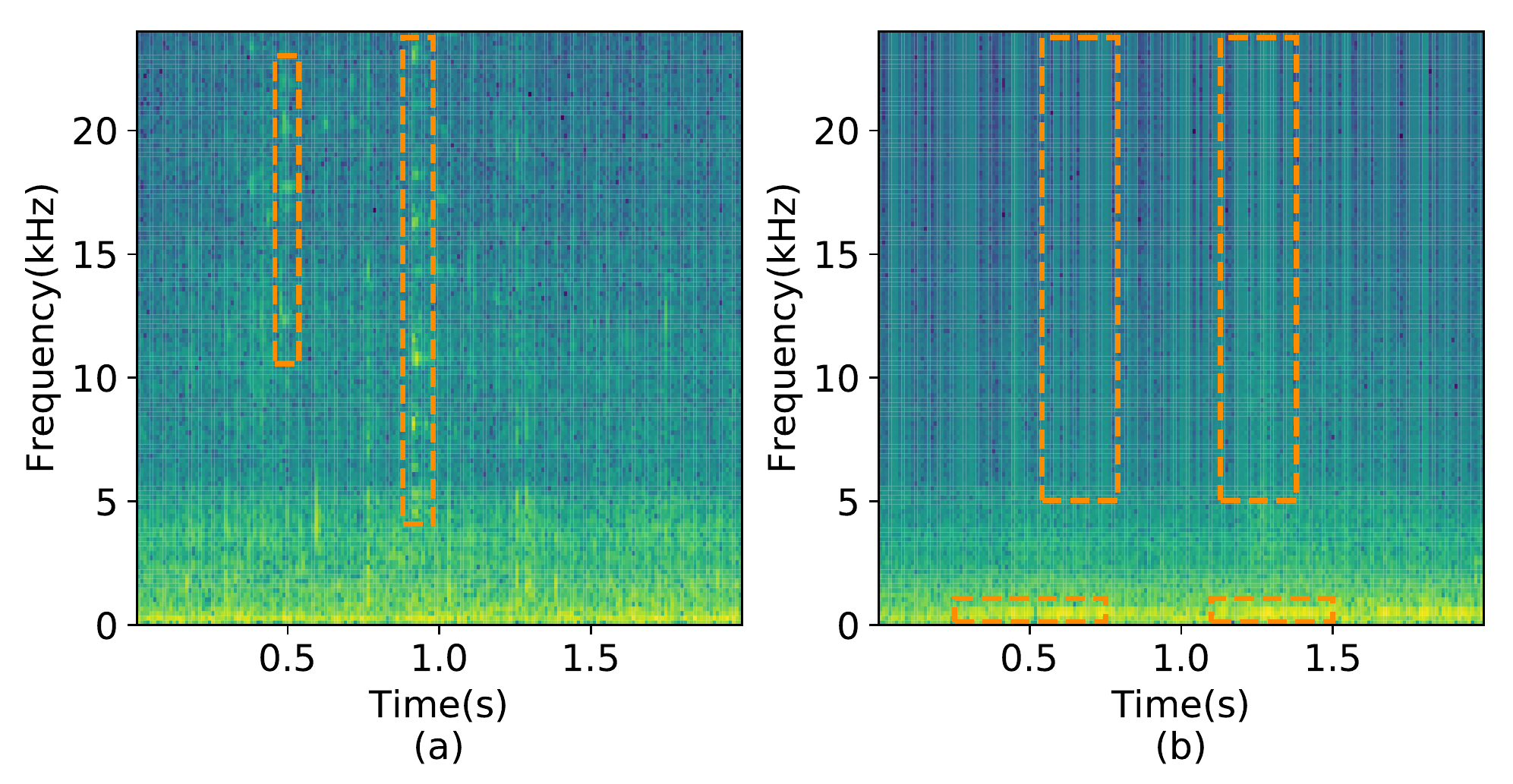}
	\caption{STFT spectra of samples recorded in (a) an airport and (b) a metro. The boxes demonstrate the acoustic cues in the scene.}
	\label{fig:STFT_cues}
\end{figure}

Accurately defining an acoustic scene is a complicated task for real-life audios. Similar to the majority of the related research, the scene recognition task in the current paper is simplified as a classification problem with predefined classes \cite{Mesaros2016TUT}. The extracted acoustic features are modeled by statistical learning methods. Since the DCASE16 competition, a large amount of collected data has supported the training of neural networks. According to the DCASE16 competition report, deep learning methods, in particular convolutional neural networks (CNNs), can perform more accurately on ASC tasks \cite{mesaros2017detection}.

Another concern is the generalization ability of models. Concretely, the classifier usually can be fit well on the training set but fails to classify testing records, particularly those from unseen cities \cite{Mesaros2017}. The balanced training data still results in biased models, yielding ambiguous results on some confusing scenes. Therefore, data augmentation approaches are useful to extend datasets and to improve the performance of classifiers. Classic methods (e.g., speed and volume perturbation \cite{ko2015audio}, pitch shift \cite{Dorfer2018}) introduce diversity increase in training dataset but cannot generate samples with new content. Generative adversarial networks (GANs), which can generate samples via a nonlinear transform, can be adopted in order to overcome the limitations of classic methods. However, controlling the number of synthetic samples to ensure performance improvements can be difficult. In the current paper, a GAN-based data augmentation scheme is proposed to iteratively extend the dataset, ensuring the improvement of the final system’s performance.

The contribution of this work is $4$-fold. First, a wavelet scale is designed to extract long-term features, which outperforms the classic Mel filters. Second, a group of neural networks, including $1D$ and $2D$ convolutional architectures for scene classifiers and GANs, are described and compared under long-term and short-term features. In particular, the adversarial city training and the discrete cosine transform (DCT)-based temporal module are proposed to increase the diversity of the models. Third, the auxiliary classifier GANs (ACGANs) are employed to directly generate raw feature maps with input conditions, instead of generating bottleneck features with individual GANs as in \cite{Mun2017}. Fourth, a novel data augmentation scheme is proposed to improve the performance by integrating ACGANs and neural network-based sample filters. This scheme demonstrates the potential of deep learning with the performance improvements achieved by utilizing iterative procedures of city-based dataset splitting, ACGAN training and sample filtering. Our system achieved the highest accuracies for DCASE19 Task 1A and delivered a state-of-the-art performance for DCASE17 Task 1. 

The remainder of the paper is organized as follows. Section \ref{sec:realted_works} reviews the previous related work. Section \ref{sec:wavelet} introduces the proposed passband wavelet filters and the algorithm employed to extract features. Section \ref{sec:clf} details the design of the classifiers. Section \ref{sec:data_aug} describes the ACGAN-based data augmentation scheme along with the detailed architectures. Section \ref{sec:exp_config} describes the dataset and the configuration of our experiments. Section \ref{sec:results} presents and discusses the results. Finally, Section \ref{sec:conclusion} concludes this work and outlines future research directions.

\section{Related Work}
\label{sec:realted_works}

The early acoustic scene classification \cite{sawhney1997situational} \cite{clarkson1998auditory} \cite{eronen2003audio} and DCASE submission systems \cite{BarchiesiAcoustic} \cite{mesaros2017detection} \cite{DCASE2018Workshop} \cite{Mesaros2019} typically consist of $2$ steps: feature extraction and classification. The first ASC system in record was built using the nearest neighbor classifier and a recurrent neural network (RNN) with FBank and other perceptual features \cite{sawhney1997situational}. In recent researches, a classic architecture is composed of FBank as the feature and a CNN as the back-end classifier \cite{Han2017}\cite{Mesaros2018}\cite{Sakashita2018}\cite{Chen2019}.

Several audio features have been exploited in ASC systems, including low level spectrum features (e.g., spectrograms \cite{Weiping2017}), auditory filter banks (e.g., FBank \cite{Mesaros2018}), cepstral features (e.g., MFCC  \cite{Dorfer2018}) and others (e.g., linear predictive coefficients \cite{BarchiesiAcoustic}). Several studies have proposed feature transform, such as non-negative matrix factorization \cite{Bisot2016} and independent component analysis \cite{Eghbal-Zadeh2016}, to derive new quantities. The constant-Q analysis is a spectral representation that fits the exponentially spaced filters \cite{brown1991calculation}. Acoustic features based on constant-Q transform (CQT) are widely used in the field of music analysis \cite{argenti2010automatic}\cite{giannoulis2011disjointess}. The deep scattering spectrum proposed in \cite{Joakim2013Deep} involves a set of evenly and geometrically-spaced wavelet filters and scattering networks, which output a group of features with multi-frequency resolutions. The CQT-based features, explored in DCASE16 \cite{Lidy2016}, 17 \cite{Weiping2017}\cite{Kun2017}\cite{amiriparian2017combined}, 18 \cite{Zeinali2018} and 19 \cite{Huang2019a}, were generally extracted using a short-term window within a range of $10$ to $50ms$ \cite{Weiping2017}\cite{Lidy2016}\cite{Zeinali2018}\cite{Bisot2017} and serve as a complementary feature \cite{Weiping2017} \cite{Zeinali2018}.

Much early research adopted generative models as classifiers, such as mixture Gaussian models \cite{chu2006scene} and hidden Markov models \cite{Ma2006Acoustic}. Constrained by limited dataset sizes, methods such as support vector machine (SVM) \cite{Elizalde2016}\cite{rakotomamonjy2014histogram} and non-negative matrix factorization \cite{Bisot2016} were also employed. Some studies attempted to infer the acoustic scene from event cues \cite{ye2015acoustic}. Since the DCASE16 challenge, the majority of the successful classifiers in the competitions have adopted deep neural networks, including multi-layer perceptron (MLP) \cite{Mun2016}, time-delay neural networks (TDNNs) \cite{Moritz2016}, CNNs \cite{Dorfer2018}\cite{Sakashita2018}, RNNs \cite{Kukanov2017} and the hybrid models \cite{Mun2017}. Some researchers built the networks involving the structures of ResNet \cite{Mingle2019}, self-attention \cite{Zhu2019} and Inception \cite{Wang2019b}. CNNs were particularly popular in the competition, with all top $10$ teams adopting CNNs as classifiers for the DCASE18 \cite{dcase2018task1aresults} and DCASE19 competition \cite{dcase2019task1aresults}. CNNs have been demonstrated to outperform other methods in the ASC task \cite{li2017comparison}\cite{chen2018deep}. This can be attributed to the little temporary dynamics but much local correlation existing in the acoustic scene spectra \cite{li2017comparison}. Two branches are employed for CNNs: $1D$ convolutional kernels along the time axis \cite{chen2018deep} and $2D$ kernels along the time-scale axis \cite{Han2017}\cite{Sakashita2018}. However, limited studies have investigated the difference between these two branches.

Inspired by the game theory, GAN is a powerful generative model with the generator and discriminator competing with each other, first introduced by \cite{goodfellow2014generative}. The generator learns from the training set and attempts to generate fake samples to fool the discriminator. Meanwhile, the discriminator attempts to classify whether the input is real or fake. Finally, a balance is reached when the fake samples from the generative model match the true distribution and the discriminator cannot distinguish between them. The GAN architecture has been developed to generate samples with labels, such as conditional GANs (CGANs) \cite{mirza2014conditional}, ACGANs \cite{odena2017conditional} and autoencoder based GANs \cite{MakhzaniSJG15}. Compared with the vanilla GAN, CGANs utilize an additional label input to generate samples from specific categories. In ACGANs, the discriminator provides the probability over the sources and the class labels. By optimizing both real/fake and label classification cost functions, the ACGANs can generate higher quality samples compared with the standard GAN \cite{odena2017conditional}. GAN and its variants have archived impressive results over a wide range of applications, including image generation \cite{denton2015deep}, style transfer \cite{isola2017image} and image super-resolution \cite{ledig2017photo}. Data augmentation using the GAN architecture has been demonstrated to improve classification performances \cite{antoniou2017data} \cite{xia2018auxiliary}. In ASC tasks, the classic data augmentation techniques, such as Mixup \cite{Huang2019a} and SpecAug \cite{Zheng2019}, have been explored. Furthermore, standard GANs with SVM have suggested improved performance with the generated bottleneck features\cite{Mun2017}.

\section{Passband Wavelet Filters}
\label{sec:wavelet}
Different acoustic scenes exhibit background sounds with various time scales. A densely sampled raw signal contains considerable meaningless fluctuations. Thus a representation of the scene information without short-term fluctuations can benefit ASC tasks. In this section, we first introduce the scalogram extracted by a set of passband wavelet filters. Then we compare FBank with the scalogram in order to investigate their differences.

\subsection{Filter definitions}
\label{subsec:passband_filters}
A scalogram is extracted with a set of passband filters. We set $\lambda_0$ as the center frequency of the mother wavelet, $Q$ as the number of filters in each octave. A dilated wavelet, indexed by $k\in \mathbb{Z}$, can be determined by its center frequency $\lambda_k$ and bandwidth $\delta_k$,
\begin{equation}
\label{eq:lambda_k}
\lambda_k=2^{k/Q}\lambda_0,
\end{equation}
\begin{equation}
\label{eq:f_k}
\delta_k=\lambda_k/Q,
\end{equation}
which follows the constant-Q property. In addition, its time interval can be described as,
\begin{equation}
\label{eq:T_k1}
T_k=\frac{1}{\delta \lambda_k}.
\end{equation}
A Gaussian-like passband filter $\hat{\phi}_k(f)$ defined in the frequency domain can be expressed as follows,
\begin{equation}
\label{eq:phi_k}
\hat{\phi}_k(f)=exp(-\frac{(\lambda_k-f)^2}{2(\delta_k)^2}),
\end{equation}
where $f$ is the frequency. This filter can be transformed into the time domain using the following, 
\begin{equation}
\phi_k(t)=\int\hat{\phi}_k(f)e^{ift}\, dt.
\end{equation}
A specific set of wavelets are determined by the upper-bound $f_h$ and the maximum length of time window $T_{max}$, which usually corresponds to the double of the maximum time interval of the wavelets. In general, the center frequency of the mother wavelet is set as $\lambda_0=f_h$ and a set of wavelets are defined by the index $k=0,...,-K+1$, where $K$ is the total number of constant-Q-spaced wavelets. The maximum time interval is defined as,
\begin{equation}
\label{eq:T_k2}
T_K=\frac{Q}{2^{-\frac{K-1}{Q}}f_h}.
\end{equation}
By setting $T_K=T_{max}/2$, $K$ can be expressed as
\begin{equation}
\label{eq:K}
K=1+Qlog_2\frac{T_{max}f_h}{2Q}.
\end{equation}

In our practice, the direct use of constant-Q wavelets to extract the scalogram results in sub-optimal performances. The length of time window $T_{max}$ limits the maximum time interval, and also determines the lowest center frequency. To cover the low frequency parts, a set of evenly-spaced passband filters are defined on the frequency domain below $2^{(-K+1)/Q}f_h$, also denoted as the wavelet for simplicity here. Their center frequency intervals are determined by the last two constant-Q wavelets. With the lower bound frequency $f_l\approx 0$, the total number of low frequency wavelets is
\begin{equation}
\label{eq:P}
P=\frac{1}{2^{1/Q}-1}.
\end{equation}
The center frequency and bandwidth of evenly-spaced passband filters, index by $k=-K,...,-K-P+1$, is
\begin{equation}
\label{eq:lambda_k2}
\lambda_k=\lambda_K-\frac{\lambda_K-f_l}{P}(K-k),
\end{equation}
\begin{equation}
\label{eq:f_k2}
\delta_k=\lambda_K/Q.
\end{equation}
In summary, given $f_h$, $f_l$ and $T_{max}$, a set of constant-Q filters are determined by Equation (\ref{eq:K}), (\ref{eq:f_k}), (\ref{eq:lambda_k}), (\ref{eq:phi_k}), and a set of evenly-spaced passband filters are determined by Equation (\ref{eq:P}), (\ref{eq:lambda_k2}), (\ref{eq:f_k2}), (\ref{eq:phi_k}). In this paper, the feature extracted by this group of filters is denoted as the scalogram.

We reformulate the equations with index $j=0,...,J-1$ ($J=K+P$) as follows,
\begin{equation}
\label{eq:summ_lambda}
\lambda_j=\left\{
\begin{aligned}
&f_l+\frac{2Q/T_{max}-f_l}{P}j\ ,\ j=0,...,P-1 \\
&\frac{2Q}{T_{max}}2^{\frac{j-P}{Q}}\ ,\ j=P,...,J-1
\end{aligned}
\right.,
\end{equation}
\begin{equation}
\label{eq:summ_delta}
\delta_j=\left\{
\begin{aligned}
&2/T_{max}\ ,\ j=0,...,P-1  \\
&\lambda_j/Q\ ,\ j=P,...,J-1
\end{aligned}
\right.,
\end{equation}
where the center frequency increases following index $j$.

The scalogram can be calculated by the convolution between the raw signal and the filters, which can be sped up using overlap-add method,
\begin{equation}
\label{eq:y(t,k)}
y(t,k)=\int_{-\delta T_k}^{\delta T_k}\phi_k(t-m)x(m)\, dm,
\end{equation}
where $y$ is the convolution result at every sample, which contains lots of fluctuations and consumes a substantial amount of computation and storage. Here the wavelet filters are directly applied to the STFT spectra following the FBank calculation. The scalogram is computed via the following steps,
\begin{enumerate}
	\item Given frame length $T_{win}$ and shift $T_{shift}$, the signal is framed and the magnitude of the STFT spectrum is calculated.
	\item Given $f_h$, $f_l$ and $T_{max}$, $J$ and $P$ are calculated. Following Equation (\ref{eq:summ_lambda}), (\ref{eq:summ_delta}), (\ref{eq:phi_k}), a group of wavelet filters are defined and their digit forms are calculated.
	\item The filters are applied to the magnitude spectrum. The sum of the power of the magnitude for each filter is taken.
	\item The logarithm of the power coefficients is taken.
\end{enumerate}
The setup calculates the convolution of the signal and the filter in a similar way to Equation (\ref{eq:y(t,k)}), but the summation of energy downsamples the point-wise feature into a frame-wise one. The scalogram has a dimension of $L\times N_{wavelet}$, where $L$ is the number of frames and $N_{wavelet}=J=K+P$ is the total number of filters.

\subsection{Comparison between FBank and scalogram}

\begin{figure}[b]
	\centering
	\includegraphics[width=0.8\linewidth]{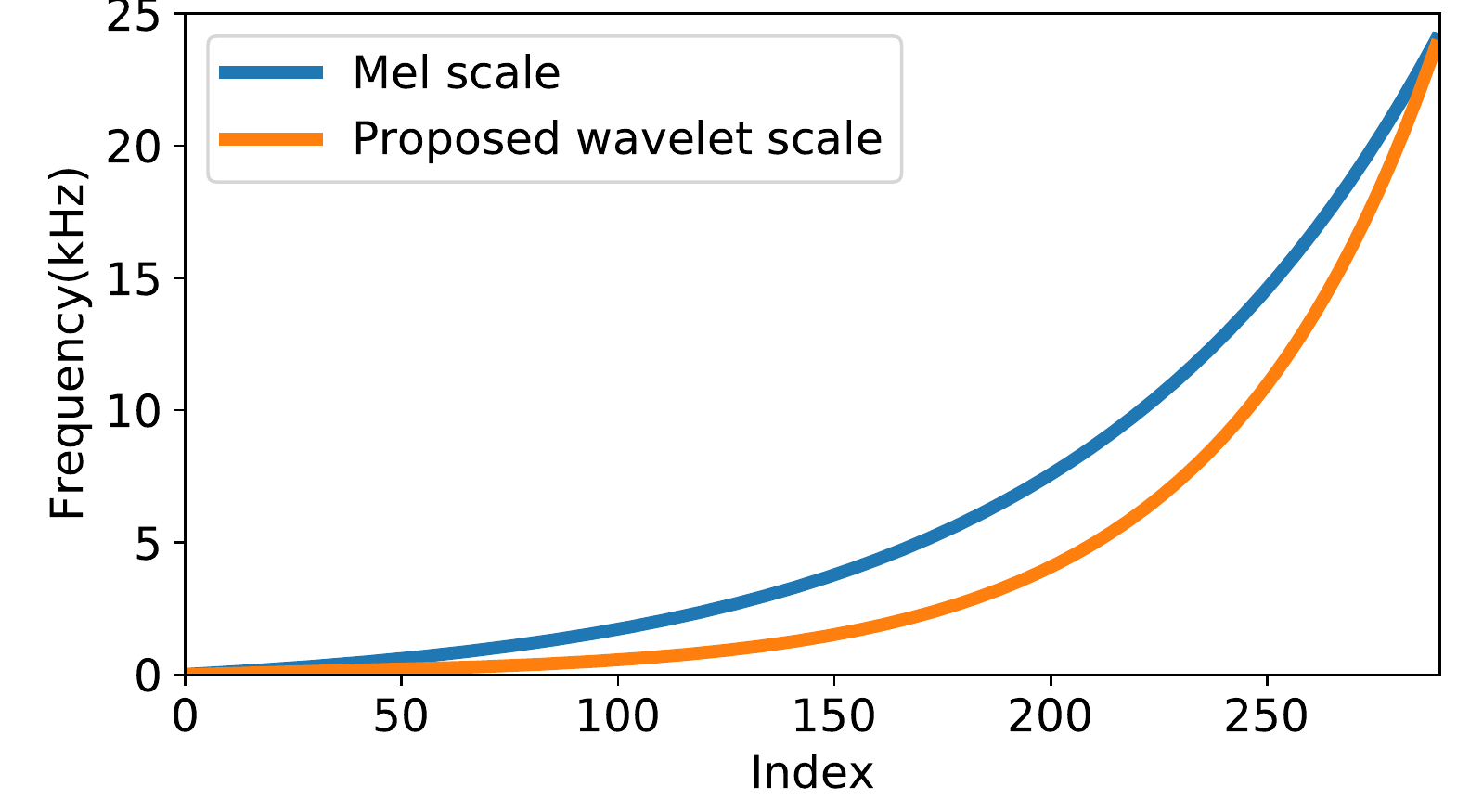}
	\caption{The Mel and wavelet scales with the same number of the filters.}
	\label{fig:melvswavelet}
\end{figure}

\begin{figure*}[!htb]
	\centering
	\centerline{\includegraphics[width=1.6\columnwidth]{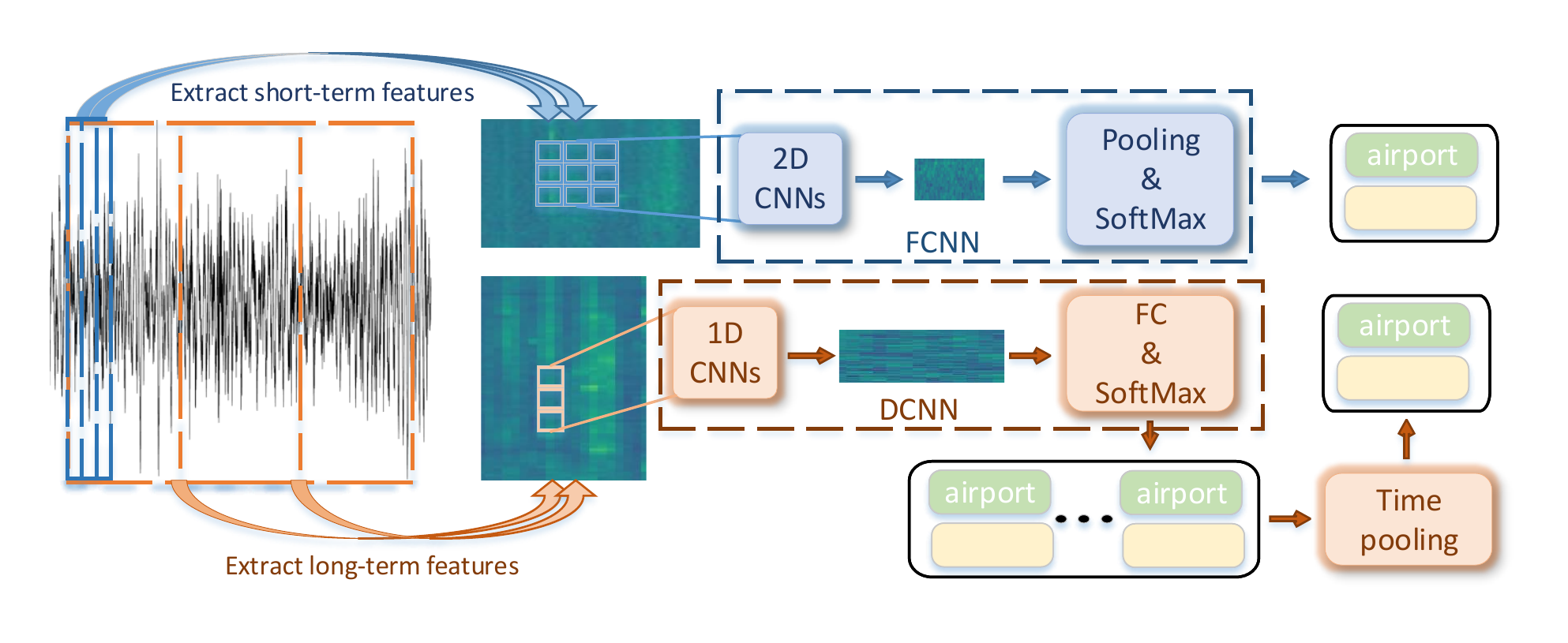}}
	\caption{The procedure of classifying an audio with FCNNs and DCNNs. The short/long-term features are extracted by the predefined filters with short/long-term windows. The FCNN processes the feature map with $2D$ convolution kernels, resulting in a small-sized high-level feature map where each pixel correlates with long-term and broadband information. The DCNN processes the feature map with $1D$ convolution kernels. Its output can be considered as a high-level broadband feature, which shares same time resolution with the input. }
	\label{fig:FDCNN}
\end{figure*}

The proposed wavelet scale is distinct to the Mel scale. The widely known Mel scale is expressed as
\begin{equation}
m(f)=781\ log_2(1+f/700),
\end{equation} 
where $m$ denotes the Mel frequency. The maximum and minimal Mel frequencies are generally defined by $m_{l}=m(f_l)$ and $m_{h}=m(f_h)$. Here, the number of Mel filters is set to $N_{Mel}$, thus the Mel frequency is evenly separated into $N_{Mel}+1$ parts. We use $m_i$ to represent the center frequencies of the $i$th filter, where $i=0,...,N_{Mel}-1$. $m_i$ is subsequently calculated as
\begin{equation}
\label{equ:mel_scale}
m_i=700\times2^{m_l/781}\times2^{\frac{(m_h-m_l)(i+1)}{781(N_{Mel}+1)}}-700.
\end{equation}
The Mel scale is actually geometrically spaced with a negative bias.

Compared with Equation (\ref{eq:summ_lambda}) and (\ref{eq:summ_delta}), the low frequency patterns of the scalogram and Mel scale are distinct. At the high frequencies, not considering the negative bias in Equation (\ref{equ:mel_scale}), the base center frequency is $700Hz$ with $2^{m_l/781}\approx0$ and $N_{Mel}\gg log_2(1+f_h/700)$, yet for the proposed wavelet filters, this value is calculated based on the maximum window length. We visualize these two scales in Figure \ref{fig:melvswavelet}, with $f_h=24kHz$, $f_l=0.5Hz$, $T_{max}=341ms$, $Q=35$ and $N_{Mel}=N_{wavelet}=290$. Under these settings, the wavelet filters focus more on the low frequency part and take larger steps at high frequencies.

\section{Convolutional Classifiers}
\label{sec:clf}

In general, CNNs achieve a better performance compared to the other architectures described in Section \ref{sec:realted_works}. As plotted in Figure \ref{fig:FDCNN}, we introduce two different neural networks: a fully convolutional neural network (FCNN) and a deep convolutional neural network (DCNN) with $2D$ and $1D$ convolution kernels for short-term and long-term features, respectively.

\subsection{FCNN}
\begin{table}[!htb]
	\caption{The FCNN Classifier. The input feature map is of size frames($L$) $\times$ channels($c_s$) $\times$ filters($n$). The notation ``$5 \times 5$ Conv(pad=$2$,stride=$2$)$\times 14c_s$-BN-ReLU'' denotes a convolutional kernel with $14c_s$ output channels and a size of $5 \times 5$, followed by batch normalization and ReLU activation.}
	\label{tab:fcnn}
	\centering
	\resizebox{0.48\textwidth}{!}{
		\begin{tabular}{cc}
			\hline  
			\textbf{Layer} & \textbf{Settings} \\
			\hline
			\hline
			Input & Fbank $L\times c_s\times n$ \\
			\hline
			Conv1 & $5 \times 5$ Conv(pad=$2$,stride=$2$)$\times 14c_s$-BN-ReLU \\
			\hline
			\multirow{2}{1.0cm}{\centering{Conv2}} & $3 \times 3$ Conv(pad=$1$,stride=$1$)$\times 14c_s$-BN-ReLU \\
			& $2 \times 2$ MaxPooling \\
			\hline
			Conv3 & $3 \times 3$ Conv(pad=$1$,stride=$1$)$\times 28c_s$-BN-ReLU    \\
			\hline
			\multirow{2}{1.0cm}{\centering{Conv4}} & $3 \times 3$ Conv(pad=$1$,stride=$1$)$\times 28c_s$-BN-ReLU    \\
			& $2 \times 2$ MaxPooling \\
			\hline
			\multirow{2}{1.0cm}{\centering{Conv5}} & $3 \times 3$ Conv(pad=$1$,stride=$1$)$\times 56c_s$-BN-ReLU    \\
			& Dropout  \\
			\hline
			\multirow{2}{1.0cm}{\centering{Conv6}} & $3 \times 3$ Conv(pad=$1$,stride=$1$)$\times 56c_s$-BN-ReLU    \\
			& Dropout  \\
			\hline
			\multirow{2}{1.0cm}{\centering{Conv7}} & $3 \times 3$ Conv(pad=$1$,stride=$1$)$\times 56c_s$-BN-ReLU    \\
			& Dropout  \\
			\hline
			\multirow{2}{1.0cm}{\centering{Conv8}} & $3 \times 3$ Conv(pad=$1$,stride=$1$)$\times 56c_s$-BN-ReLU    \\
			& $2 \times 2$ MaxPooling \\
			\hline
			\multirow{2}{1.0cm}{\centering{Conv9}} & $3 \times 3$ Conv(pad=$0$,stride=$1$)$\times 128c_s$-BN-ReLU    \\
			& Dropout  \\
			\hline
			\multirow{2}{1.0cm}{\centering{Conv10}} & $3 \times 3$ Conv(pad=$0$,stride=$1$)$\times 128c_s$-BN-ReLU    \\
			& Dropout  \\
			\hline
			\multirow{2}{1.0cm}{\centering{Pooling}} & $1 \times 1$ Conv(pad=$0$,stride=$1$)$\times c_s^{scene}$-BN-ReLU    \\
			& GlobalAveragePooling    \\
			\hline
			Output & $c_t^{scene}$-way SoftMax  \\
			\hline
	\end{tabular}}
\end{table}

The short-term features yield a strong local correlation along both the time and scale axis. To model this characteristic, a FCNN architecture similar to the VGG-style network \cite{simonyan2015very} proposed in \cite{Dorfer2018} is adopted here (Table \ref{tab:fcnn}). The network consists of repeatedly stacked layers, each including a convolutional operation with small-sized $2D$ kernels, batch normalization and ReLU activation. The maximum pooling and dropout layers are interleaved between some convolutional layers. A $1\times1$ kernel in the last convolutional layer reduces the channel to $c_t^{scene}$, with the feature map averaged over each channel. The whole network is fully convolutional, which reduces the number of parameters and makes the learned transform both time and scale invariant. During the test, the scene is chosen corresponding to the maximum log probability.

\subsection{DCNN}

\begin{table}[!htb]
	\caption{The DCNN Classifier. The input feature map is of size frames($L$) $\times$ channels($c_s$) $\times$ filters($n$). The notation 
		``$3$ Conv(pad=$0$,stride=$1$)-$2c_s$-BN-ReLU'' denotes a $1D$ convolutional kernel with $2c_s$ output channels and a size of $3$, followed by batch normalization and ReLU activation.}
	\label{tab:dcnn}
	\centering
	\resizebox{0.48\textwidth}{!}{
		\begin{tabular}{cc}
			\hline
			\textbf{Layer} & \textbf{Settings} \\
			\hline\hline
			Input & Scalogram $L\times c_s\times n$ \\
			\hline
			\multirow{2}{1.0cm}{\centering{Conv1}} & $3$ Conv(pad=$0$,stride=$1$)$\times2c_s$-BN-ReLU             \\
			& $2$ Pooling(pad=$1$,stride=$2$)              \\
			\hline
			\multirow{2}{1.0cm}{\centering{Conv2}}  & $3$ Conv(pad=$0$,stride=$1$)$\times4c_s$-BN-ReLU             \\
			& $2$ Pooling(pad=$0$,stride=$2$)-Dropout              \\
			\hline
			\multirow{2}{1.0cm}{\centering{Conv3}} & $3$ Conv(pad=$0$,stride=$1$)$\times8c_s$-BN-ReLU             \\
			& $2$ Pooling(pad=$0$,stride=$2$)              \\
			\hline
			\multirow{2}{1.0cm}{\centering{Conv4}} & $3$ Conv(pad=$0$,stride=$1$)$\times16c_s$-BN-ReLU             \\
			& $2$ Pooling(pad=$0$,stride=$2$)-Dropout              \\
			\hline
			Concat & Concatenate and flatten input as well as Conv's output \\ 
			\hline
			FC1 & Linear ($h$ units)-BN-ReLU-Dropout \\ 
			\hline
			FC2 & Linear ($h$ units)-BN-ReLU-Dropout \\ 
			\hline
			FC3 & Linear ($h$ units)-BN-ReLU \\ 
			\hline
			Output & $c_t^{scene}$-way SoftMax \\ 
			\hline
	\end{tabular}}
\end{table}

The long-term features exhibit a weak correlation along time. Thus, the DCNN serves as the back-end classifier (Table \ref{tab:dcnn}), which employs small $1D$ convolutional kernels to learn high-level scale patterns. The network is first stacked by $4$ repeated layers, each including a $1D$ convolutional operator along the scale axis, batch normalization, ReLU activation and average pooling. The channel number of convolutional kernels doubles after each layer. Note that before being fed into fully connected layers, the output of each convolutional layer, as well as the input feature, is concatenated into one super vector for each frame. The convolutional kernel generates high-level patterns to increase data discrimination. Both the input feature map and the output of the convolutional kernels benefit the system performance. During the test, the scene is chosen corresponding to the maximum sum of the frame-wise log probability. 

Two additional techniques employed for DCNNs, adversarial city training and a temporal module based on the DCT, are introduced in order to improve the performance.

\subsubsection{Adversarial city training}
The scene classifier cannot generalize well on records from unseen cities. Thus, an adversary training branch, composed of a gradient reverse layer (GRL) \cite{Ganin2015Domain} and a $2$-layer fully-connected classifier, is employed following the ``Concat'' layer in Table \ref{tab:dcnn} to perform adversarial city training. This branch classifies the audios into their recorded cities with cross entropy city loss $L(clf_{city}(x_i),y_i^{city})$ where $clf_{city}(x_i)$, $x_i$, $y_i^{city}$ represent the output of the city branch, the feature and the city label of sample $i$, respectively. The overall loss $L_{adv}$ can be written as
\begin{equation}
\begin{aligned}
\label{eq:l_adv}
L_{adv}=& L(clf_{scene}(x_i),y_i^{scene}) \\
	&-\gamma_{adv}L(clf_{city}(x_i),y_i^{city}),
\end{aligned}
\end{equation}
where $clf_{scene}(x_i)$, $y_i^{scene}$ represent the output of the main scene classifier and the scene label of sample $i$, respectively. $\gamma_{adv}$ balances the scene and the city loss. During training, the minimax game is conducted by the GRL, which introduces the negative part of $L_{adv}$ by gradient reverse. In detail, the parameters of the main scene classifier including its shallow convolutional kernels are optimized in order to minimize $L_{adv}$ (i.e., to minimize the scene loss  $L(clf_{scene}(x_i),y_i^{scene})$ and to maximize the city loss $L(clf_{city}(x_i),y_i^{city})$). The parameters of the adversary training branch are optimized to minimize $L(clf_{city}(x_i),y_i^{city})$. This minimax competition will first increase the discrimination of city branch and the city-invariance of the features generated by convolutional kernels from the scene classifier. Eventually it will occur to the point where the output of the convolutional kernels is similar for the same scene over different city domains.

\subsubsection{DCT-based temporal module}

\begin{figure}[!htb]
	\centering
	\centerline{\includegraphics[width=1.0\columnwidth]{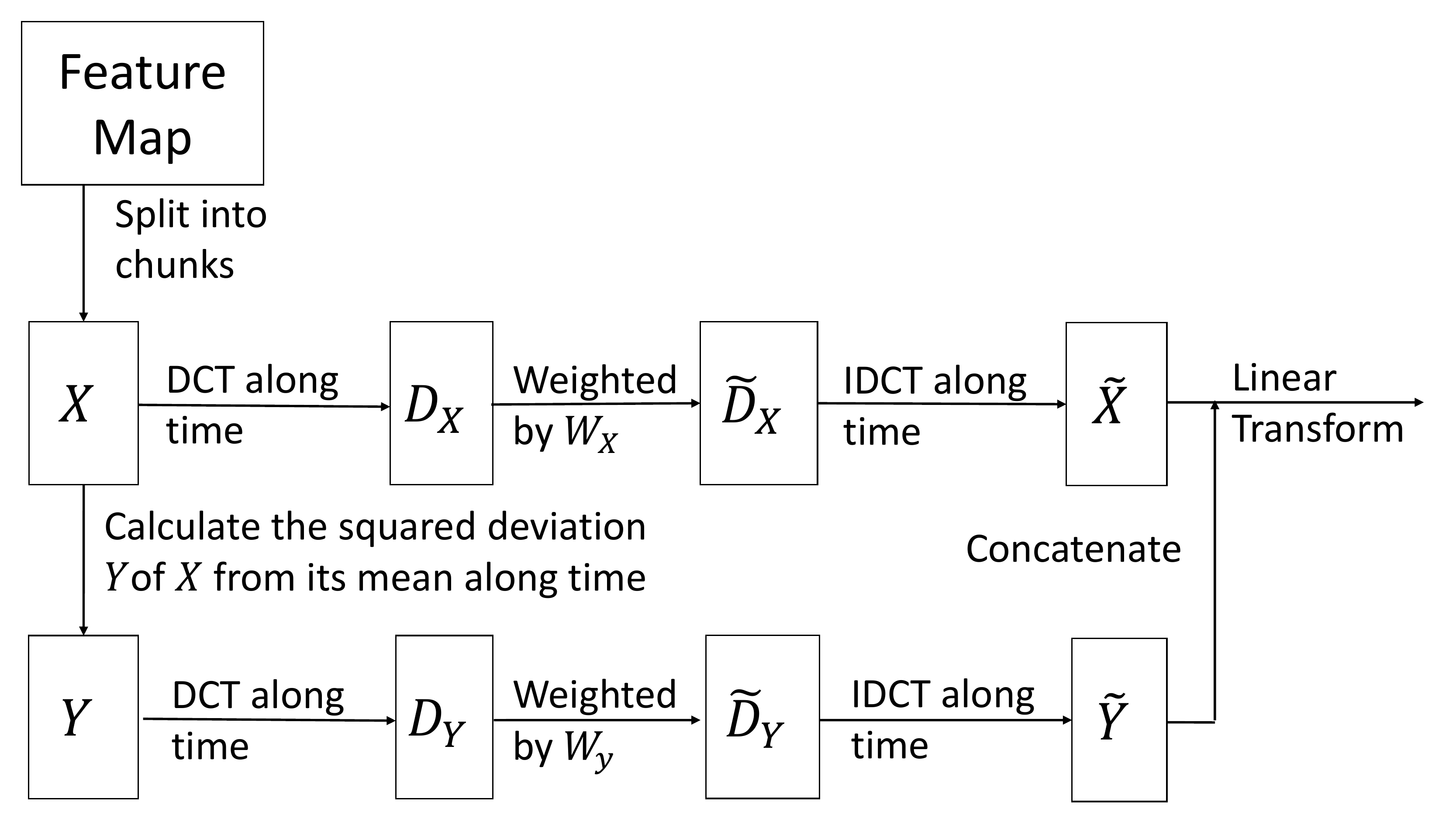}}
	\caption{The DCT-based temporal module.}
	\label{fig:dct}
\end{figure}

The DCT-based temporal module utilizes information across time with learnable weight matrices  (Figure \ref{fig:dct}) \cite{chen2019audio}. The input feature map is first split into non-overlapping chunks $X\in R^{T\times N}$ along with the second order statistical feature $Y\in R^{T\times N}$, where $T$ is the frame number in a chunk and $N$ is the feature dimension. The chunks $X$ and $Y$ are transformed into the DCT domain and weighted by the learnable weights $W_X$ and $W_Y$. After the inverse DCT, the outputs $\tilde{X}$ and $\tilde{Y}$ are concatenated along the feature dimensions. They are subsequently linearly transformed into a feature with the same size of the original input.

This module is inspired by noise reduction approaches in image processing. Here, the weights can be viewed as attention-like coefficients, which strengthen and weaken the target feature map bins. Moreover, the module can be inserted without changing the original network topology as its input and output share the same size. 

\section{Data augmentation scheme}
\label{sec:data_aug}

Although ASC systems are able to accurately classify most training samples, they may suffer from inferring some test records, particularly those lying on the margin of classification planes, collected in unseen cities and belonging to confusing scenes. To improve the generalization performance, we apply a GAN-based data augmentation scheme. Distinct from \cite{Mun2017}, which built individual GANs for each acoustic scene and generated bottleneck features, our proposed technique performs conditional acoustic feature synthesis with ACGANs to directly generate different spectrum samples from one unified model. In addition, a novel iterative scheme combining ACGANs and a deep-learning-based sample filter is deployed to yield stable performance improvements.

\subsection{ACGAN}
The GAN plays a minimax game where the discriminator is optimized to discriminate real/fake samples and the generator attempts to fool the discriminator. More specifically, for ACGAN with the condition, a generator learns to generate acoustic features with randomly sampled noise and a one-hot condition as input. The main branch of the discriminator is trained to discriminate whether the input sample is real or fake. On the other hand, an additional branch classifies the input into scenes. In formulation, the discriminator network $D$ with parameters $\theta_D$ predicts the posterior source probability of the real samples $s_i$ and the fake ones $\tilde{s}_i$, which are generated by the generator $G$ with parameter $\theta_G$, 
\begin{equation}
P(Real | x_i;\theta_D)=D_{source}(x_i),
\end{equation}
\begin{equation}
P(Fake | x_i;\theta_D)=1-D_{source}(x_i),
\end{equation}
where $x_i$ is $s_i$ or $\tilde{s}_i$.
The $G/D$ is optimized to maximum/minimize the following real/fake loss,
\begin{equation}
\begin{aligned}
	&L_{source}=-\sum_{i}\{logP(Real | s_i;\theta_D)+logP(Fake | \tilde{s}_i;\theta_D)\} \\
	&=-\sum_{i}\{logD_{source}(s_i)+log[1-D_{source}(G(z,y_i))]\}, 
\end{aligned}
\end{equation}
where $z$ and $y_i$ are the noise and the scene condition, respectively.

To make the generated samples not only look real but belong to the target scene, we perform an additional branch located at $D$ for the scene classification. The auxiliary classification loss is formulated as
\begin{equation}
\begin{aligned}
L_{scene}=-\sum_{i}\sum_{a\in A}{\mathbb I}_{[a=y_i]}\{&log(D_{scene}(x_i)) \\
+&log(D_{scene}(G(z,y_i)))\},
\end{aligned}
\end{equation}
where $A$ is a collection of scene classes, $\mathbb{I}$ is the indicator function, $y_i$ is the scene label of $x_i$. $G$ and $D$ are separately trained to optimize the multi-task loss as follows,
\begin{equation}
\label{eq:theta_D}
\hat{\theta}_D=\mathop{\arg\min}_{\theta_D} \ L_{source}+\gamma_{aux} L_{scene},
\end{equation}
\begin{equation}
\label{eq:theta_G}
\hat{\theta}_G=\mathop{\arg\min}_{\theta_G} \ -L_{source}+\gamma_{aux} L_{scene},
\end{equation}
where $\gamma_{aux}$ controls the ratio between the source and scene classification loss.

\subsection{Augmentation scheme}

\begin{figure}[!htb]
	\centering
	\centerline{\includegraphics[width=1.0\columnwidth]{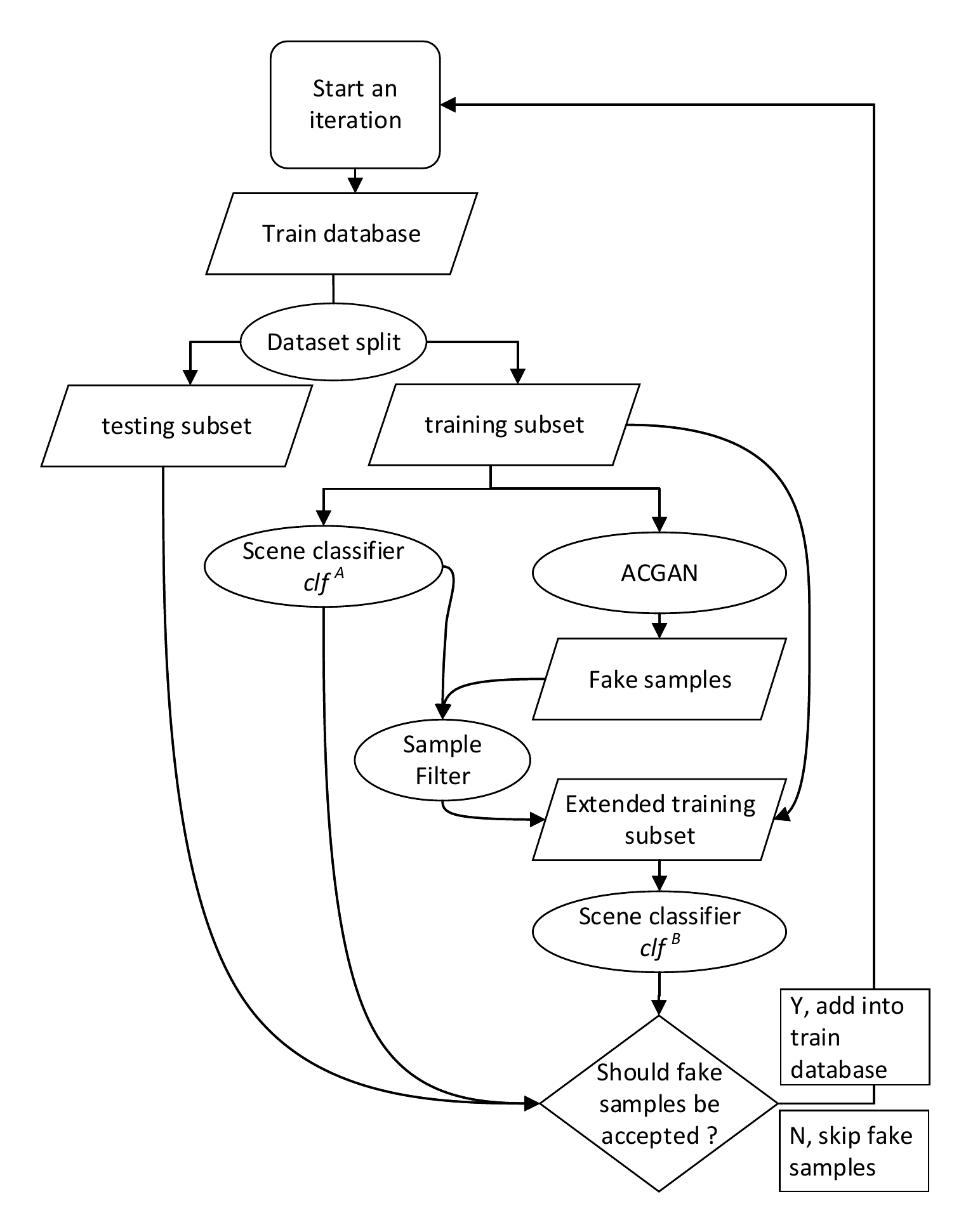}}
	\caption{The scheme of data augmentation to iteratively extend the database.}
	\label{fig:scheme}
\end{figure}

The generated fake samples are included into the dataset with verification processes to ensure performance improvements. An iterative procedure is conducted to test whether the generated samples are indeed useful for the performance. As shown in Figure \ref{fig:scheme}, at the $k$th iteration, the whole training database is first split into training and testing subset. A scene classifier $clf^A$ is trained using the training subset with early-stopping controlled by the testing subset. Meanwhile, an ACGAN is trained on the training subset. To guarantee that the generated samples benefit both the current and final classifiers, a simple sample filter is deployed by piping the fake samples through classifier $clf^A$ , and keeping only those lying on the classification margin. A new scene classifier $clf^B$ is further trained on the extended dataset, which contains current fake samples filtered by the sample filter and the original subset. The $clf^A$ and $clf^B$'s performance test is conducted on the testing subset. If $clf^B$ achieves a higher classification accuracy, these samples are accepted and merged into the whole database, otherwise they are abandoned. A final classifier is built on the whole database, including both real and accumulated fake samples. We address the dataset splitting and the sample filter as follows.

We apply $3$ methods to split the dataset in the experiments. The simplest one is the fixed splitting, which splits the whole database using a fixed random seed and exchanges the training and testing sets during the iterations. The second method employs different random seeds to split those datasets at each iteration. The third randomly splits samples according to city information. 

Depending on the characteristics of the input features, we can generate frame-wise and segment-wise samples with different generator architectures. To sample a scene, $N_{sample}$ and $C_{sample}$ are set as the maximum number of samples and sample time, respectively. More specifically, the sampling procedure stops if either of these two conditions is met. A sample filter determines whether the sample lies on the classification borderline. For the $C$-class classification, the borderline is set as the output probability equal to $1/C$ with margin $m_p$. The algorithm used to filter frame-wisely generated samples is described in Algorithm \ref{alg:alg1}. The segment-wisely generated samples are processed similarly, except that the operation is performed on the segment level.  

\begin{algorithm}[t]
	\caption{The sample filter for frame-wisely generated features}
	\hspace*{0.02in} {$clf^A$:}
	Scene classifier trained on the training subset\\
	\hspace*{0.02in} {$\{G\}$:}
	A set of generators from different training epochs\\
	\hspace*{0.02in}{$N_{sample}$:}
	maximum number of samples for each scene\\
	\hspace*{0.02in}{$T_{sample}$:}
	maximum number of sample times\\
	\hspace*{0.02in}{$C$:}
	The number of scene classes\\
	\hspace*{0.02in}{}
	The feature is of size $L\times c_s\times n$
	\begin{algorithmic}[1]
		\State Maximum number of frames sampled from each $G$ $N_{sample}^{frame}=L\times N_{sample}/\#\{G\}$
		\State An empty candidate database $\Lambda$
		\For{each scene label $z$, each $G$ in $\{G\}$}
		\State Current state $n_{sample}^{frame}:=0$, $t_{sample}:=0$
			\While{$t_{sample} < T_{sample}$ and $n_{sample}^{frame} < N_{sample}^{frame}$} 
			\State Generate $N_{sample}^{frame}$ frames from $G$, each frame $f \in\mathbb{R}^{c_s\times n}$
			\For{each $f$ in $N_{sample}^{frame}$ frames}
				\If{$1/C-m_p<clf^A(f)[z]<1/C+m_p$} 
				\State $\Lambda:=\Lambda \cup f$
				\State $n_{sample}^{frame}:=n_{sample}^{frame}+1$
				\EndIf 
			\EndFor
			\State $t_{sample}:=t_{sample}+1$
			\EndWhile
		\EndFor
		\State The number of generated samples $N_{gen}=\lfloor\#\Lambda/L\rfloor$
		\State Reshape $\Lambda$ into a size of $N_{gen}\times L\times c_s\times n$
		\State \Return $\Lambda$
	\end{algorithmic}
	\label{alg:alg1}
\end{algorithm}

\subsection{Network architecture}
Two sets of ACGAN architectures are designed to generate segment-wise and frame-wise samples, corresponding to short-term and long-term features respectively (Table \ref{tab:seg_gen} and \ref{tab:frame_gen}). The discriminators are simplified versions of the aforementioned DCNNs and FCNNs. The generators are mirror images of the discriminators. To generate long-term features, $1D$ convolutional kernels are applied throughout, which means that the temporal dynamics is not considered. To avoid the training collapse of generators, the leaky ReLU replaces the original sparse ReLU \cite{Radford2015Unsupervised}. Meanwhile, the number of convolution layers is reduced to half. The randomness of the generator is from the noise multiplying the embeddings indicated by the one-hot scene condition.

\begin{table*}
	\begin{minipage}{0.48\linewidth}
		\centering
		\caption{The segment-wise ACGAN with segment-wise input and output.}
		\label{tab:seg_gen}
		\begin{tabular}{c|c} 
			\hline
			\textbf{Layer} & \textbf{Settings} \\
			\hline\hline
			Input & One-hot scene Label $c_t^{scene}$ \& Noise $h_{emd}$ \\
			\hline
			Embedding & Embeddings $h_{emd}$ $\times$ Noise  \\
			\hline
			FC1 & Linear($h$ units)-BN-LeakyReLU \\
			\hline
			FC2 & Linear($h_{G2}$ units)-BN-LeakyReLU \\
			\hline
			Reshape & Reshape to bottleneck feature $32c_G\times L_{bn}\times n_{bn}$ \\
			\hline
			DeConv1 &  $5\times5$ DeConv(pad=2,stride=4)$\times8c_G$-BN-LeakyReLU \\
			\hline
			DeConv2 &  $5\times5$ DeConv(pad=2,stride=4)$\times c_G$-BN-LeakyReLU \\
			\hline
			FC3 &  Linear($L$ units along time axis) \\
			\hline
			FC4 & Linear($n$ units) \\
			\hline
			Output & Permute the dimensions to size $L\times c_G\times n$ \\
			\hline\hline
			Input & Feature $L\times c_G\times n$ \\
			\hline
			\multirow{2}{1.0cm}{\centering{Conv1}} & $5\times5$ Conv(pad=$2$,stride=$1$)$\times8c_s$-BN-LeakyReLU             \\
			& $2$ Pooling(stride=$4$)-Dropout2D             \\
			\hline
			\multirow{2}{1.0cm}{\centering{Conv2}} & $5\times5$ Conv(pad=$2$,stride=$1$)$\times32c_s$-BN-LeakyReLU             \\
			& $2$ Pooling(stride=$4$)-Dropout2D             \\
			\hline
			Reshape & Reshape each segment into a vector \\
			\hline
			FC1 & Linear($h$ units)-BN-LeakyReLU \\
			\hline
			Output & Linear($c_{t}^{scene}$ units)-SoftMax \& Linear($1$ unit) Sigmoid \\
			\hline
		\end{tabular} 
	\end{minipage}
	\begin{minipage}{0.48\linewidth}  
		\centering
		\caption{ The frame-wise ACGAN with frame-wise input and output.}
		\label{tab:frame_gen}
		\begin{tabular}{c|c} 
			\hline
			\textbf{Layer} & \textbf{Settings} \\
			\hline\hline
			Input & One-hot scene Label $L\times c_t^{scene}$ \& Noise $L\times h_{emd}$ \\
			\hline
			Embedding & Embeddings $h_{emd}$ $\times$ Noise  \\
			\hline
			Reshape & $L\times16c_s\times n_{bn}$ \\
			\hline
			DeConv1 &  $5$ DeConv(pad=0,stride=4)$\times8c_s$-BN-LeakyReLU \\
			\hline
			DeConv2 &  $5$ DeConv(pad=0,stride=4)$\times c_s$-BN-LeakyReLU \\
			\hline
			Output1 & Linear($n$ units) \\
			\hline\hline
			Input & Feature $L\times c_s\times n$ \\
			\hline
			\multirow{2}{1.0cm}{\centering{Conv1}} & $5$ Conv(pad=$0$,stride=$1$)$\times4c_s$-BN-LeakyReLU             \\
			& $2$ Pooling(stride=$4$)-Dropout2D             \\
			\hline
			\multirow{2}{1.0cm}{\centering{Conv2}} & $5$ Conv(pad=$0$,stride=$1$)$\times16c_s$-BN-LeakyReLU             \\
			& $2$ Pooling(stride=$4$)-Dropout2D             \\
			\hline
			Concat & Concatenate and flatten input as well as Conv's output \\
			\hline
			FC1 & Linear($h$ units)-BN-LeakyReLU \\
			\hline
			Output & Linear($c_{t}^{scene}$ units)-SoftMax \& Linear($1$ unit)-Sigmoid \\
			\hline
		\end{tabular} 
	\end{minipage}
\end{table*}

\section{Experimental Configuration}
\label{sec:exp_config}

\subsection{Dataset}

The experiments were conducted on the DCASE19 Task 1A dataset, which includes both the development and evaluation set. As the DCASE committee does not intend to release the ground truth of the evaluation set, we directly evaluated the systems on the officially provided fold-$1$ setup. In terms of the published competition results, the accuracies of the fold-$1$ evaluation procedure are sufficient for comparisons. Approximately $13\%$ of the fold-$1$ training set was randomly selected and reserved as the validation set for hyperparameter fine-tuning and early stopping. The systems were evaluated using the fold-$1$ evaluation set. Key experiments were also performed on the DCASE17 Task 1 dataset for fair comparisons with the SVM-GAN-based data augmentation framework \cite{Mun2017}. Overall accuracy is used as the evaluation metric, calculated as the percentage of correctly classified segments out of the total number. This measure is slightly different from the average of the class-wise accuracy when the number of samples is imbalanced among different classes. Experimental results demonstrate that the overall accuracy is slightly lower (by approx. $0.10\%$) than the average class-wise accuracy on the DCASE19 fold-$1$ dataset. 

\begin{table}[t]
	\renewcommand{\arraystretch}{1.3}
	\caption{Information of the fold-$1$ setup on the DCASE19 Task 1A Dataset}
	\label{tab:DCASE19}
	\centering
	\begin{tabular}{c|c|c}
		\hline
		\textbf{Attribute} & Development set & Evaluation set \\
		\hline
		Total number & $9185$ & $4185$ \\
		\hline
		Total runtime(h) & $25.5$ & $11.6$ \\
		\hline
		\multirow{2}{*}{WAV format}  & \multicolumn{2}{c}{Dual channel, $48kHz$ sample rate} \\
		 & \multicolumn{2}{c}{$24$-bit resolution, $10$-second duration} \\
		\hline
		\multirow{5}{*}{Acoustic scenes} & \multicolumn{2}{c}{Airport, Indoor shopping mall,} \\
		& \multicolumn{2}{c}{Metro station, Pedestrian street, Public square} \\
		& \multicolumn{2}{c}{Street with medium level of traffic,} \\
		& \multicolumn{2}{c}{Traveling by a tram, Traveling by a bus} \\
		& \multicolumn{2}{c}{Traveling by an underground metro, Urban park} \\
		\hline
	\end{tabular}
\end{table}

The DCASE19 audios were recorded with microphones that mimic headphones, with recordings sounding similar to those listened to in ears \cite{Mesaros2019}\cite{Mesaros2018}. The dataset contains $10$ acoustic scene classes. For the fold-$1$ setup, $9$ European cities are included in both the training and evaluation sets. In addition, the evaluation set contains a new city that is unseen for the trained systems. A detailed description of the datasets is reported in Table \ref{tab:DCASE19}. The official baseline achieved an accuracy of $62.5\%$ on the fold-$1$ evaluation set \cite{Mesaros2019}.

The DCASE17 dataset contains $15$ scene classes. The official baseline achieved $61.0\%$ on the evaluation set \cite{Mesaros2017}.

\subsection{Features}
\label{sec:exp_feat}
Table \ref{tab:network_params} reports the detailed settings of the scalogram and FBank. The features were extracted on the STFT spectra with predefined filters. Global normalization was performed according to the mean and standard derivation calculated on the training set. The features were extracted separately from the dual channels and concatenated along the channel axis to form a $3D$ feature map. The \textit{left-right feature} was named after being from the left and right audio channels. Furthermore, the \textit{ave-diff feature} was extracted from the mean and differential signals.

For scalogram, given $f_h$, $f_l$ and $T_{max}$, $K$ the number of constant-Q-spaced wavelets was set to $241$ (Equation (\ref{eq:K})). The number of evenly-space passband filter covering the low frequency part was $49$ (Equation (\ref{eq:P})).

We followed the procedure in \cite{Young2006The} to define the Mel filters. The vanilla FBank of a $10$-second stereo audio had dimensions of $500\times2\times128$. The delta and delta-delta coefficients were calculated and stacked along the channel axis to form a feature map of size $500\times6\times128$.

The short-term FBank exhibited a higher time resolution compared to the long-term scalogram. In order to illustrate the long-term and short-term properties, the cosine similarity was averaged every $2$ frames with randomly chosen audio samples. The dimensions of the features were unified with principle component analysis. The similarity values for Fbank and the scalogram were $0.80$ and $0.54$, respectively. This indicates that the FBank extracted by a $40ms$ window exhibits a greater temporal correlation compared to the scalogram.

\begin{table}[tb]
	\renewcommand{\arraystretch}{1.3}
	\caption{The detailed settings for features, scene classifiers and GANs.}
	\label{tab:network_params}
	\centering
	\begin{tabular}{c|c|c}
		\hline
		\multicolumn{3}{c}{\textbf{Features}} \\ \hline
		\multirow{6}{*}{Scalogram} & STFT window size & $512ms$ \\ \cline{2-3}
		& STFT shift size & $171ms$ \\ \cline{2-3}
		& $f_h$ & $24000Hz$ \\ \cline{2-3}
		& $f_l$ & $0.5Hz$ \\ \cline{2-3}
		& $Q$ & $35$ \\ \cline{2-3}
		& $T_{max}$ & $341ms$ \\ \cline{2-3}
		& Feature size & $58\times 2 \times 290$ \\ \hline
		\multirow{5}{*}{FBank} & STFT window size & $40ms$ \\ \cline{2-3}
		& STFT shift size & $20ms$ \\ \cline{2-3}
		& $f_h$ & $24000Hz$ \\ \cline{2-3}
		& $f_l$ & $0$ \\ \cline{2-3}
		& Feature size & $500\times6\times128$ \\ \hline
		\multicolumn{3}{c}{\textbf{Networks}} \\ \hline
		\multirow{2}{*}{Shared hyperparameter} & $c_t^{scene}$ & $10$ \\ \cline{2-3}
		& $h$ & $1024$ \\ \hline
		\multirow{3}{*}{}DCNN (Table \ref{tab:dcnn}) & $L\times c_s\times n$ & $58\times2\times290$ \\ \cline{2-3}
		\& frame-wise ACGAN & $h_{emb}$ & $576$ \\ \cline{2-3}
		(Table \ref{tab:frame_gen}) & $n_{bn}$ & $18$ \\ \hline
		\multirow{6}{*}{} & $L\times c_s\times n$ & $500\times6\times128$ \\ \cline{2-3}
		FCNN (Table \ref{tab:fcnn}) & $c_G$ & $2$ \\ \cline{2-3}
		\& segment-wise ACGAN & $h_{emb}$ & $500$ \\ \cline{2-3}
		(Table \ref{tab:seg_gen}) & $h_{G2}$ & $15872$ \\ \cline{2-3}
		& $L_{bn}$ & $31$ \\ \cline{2-3}
		& $n_{bn}$ & $8$ \\ \hline
		\multirow{4}{*}{Training settings} & Batch size & $128$ \\ \cline{2-3}
		& Dropout rate & $0.5$ \\ \cline{2-3}
		& Initial learning rate & $10^{-3}$ \\ \cline{2-3}
		& Maximum epochs & $200$ \\ \hline
		\multirow{4}{*}{Optional module} & $\gamma_{adv}$ (adversarial & \multirow{2}{*}{$0.1$} \\
		& city training) & \\ \cline{2-3}
		& Chunk size (DCT-based & \multirow{2}{*}{$18$} \\
		& temporal module) & \\ \cline{2-3}
		\hline
		\multicolumn{3}{c}{\textbf{Data augmentation}} \\ \hline
		\multirow{3}{*}{GAN training settings} & Maximum epochs & $50$ \\ \cline{2-3}
		& $G/D$ training times & $3/1$ \\ \cline{2-3}
		& $\gamma_{aux}$ & $0.2$ \\ \hline
		\multirow{8}{*}{Algorithm \ref{alg:alg1}} & \multirow{2}{*}{$\{G\}$} & {$35, 40, 45, 50th$}   \\ 
		& & iteration models \\ \cline{2-3}
		& \multirow{2}{*}{$N_{sample}$} & $8$ (frame-wise) \\
		& & $6$ (segment-wise) \\ \cline{2-3}
		& \multirow{2}{*}{$T_{sample}$} & $10$ (frame-wise) \\
		& & $8$ (segment-wise) \\ \cline{2-3}
		& $C$ & $10$ \\ \cline{2-3}
		& $m_p$ & $0.03$ \\ \hline
	\end{tabular}
\end{table}

\subsection{Networks}

The detailed hyperparameters for the networks in Table \ref{tab:fcnn}-\ref{tab:frame_gen} are listed in Table \ref{tab:network_params}. We designed the FCNNs and the segment-wise ACGANs for FBank, the DCNNs and frame-wise ACGANs for the scalogram.

The networks were trained and evaluated using PyTorch toolkit \cite{paszke2017automatic}. Adam was used to update the parameters \cite{Kingma2014Adam}. The batch normalization was performed in $1D$ and $2D$ for each channel in DCNNs and FCNNs. A slow version of early stopping was employed for all models trained on the whole training database. More specifically, the learning rate was reduced by $50\%$ following $5$ continuous epochs without a reduction of validation loss. The training would be terminated if the validation loss failed to decrease over $15$ continuous epochs. The models trained on the training subset during data augmentation, containing $clf^A$ and $clf^B$, were optimized via a fast early-stopping. The learning rate was reduced to $50\%$ following $3$ continuous epochs, the training was terminated after $6$ continuous epochs based on the testing subset.

\subsection{Data augmentation}

The data augmentation configuration is reported in Table \ref{tab:network_params}. The feature database used for the training of the final classifiers included real samples extracted from raw signals as well as generated fake samples. The GAN-based data augmentation was carried out according to Figure \ref{fig:scheme}, with maximum $10$ iterations. The generation and filtering of samples was performed via Algorithm \ref{alg:alg1}. If no sample successfully passed the sample filter, the current iteration was rejected. If scene classifier $clf^B$ performed worse than $clf^A$, the iteration was also marked as a rejection, otherwise it was accepted. The iterative data augmentation scheme would be terminated if rejected over $3$ continuous iterations.

\subsection{System fusion}

Due to the limited dataset size and training stability, the performance of each system was represented as the average voting of the system trained by $3$ random seeds. This was able to partially remove randomness and allowed for fair comparisons.

In the experiments, our base classifiers were able to achieve high accuracies. Complex fusion methods, such as the random forest and SVMs, resulted in severe overfitting. Average voting served as a simple yet flexible approach by averaging the segment-wise log-probability of all systems. For DCNNs, the segment-wise log-probability is the mean of the frame-wise ones. Note that for our DCASE19 challenge submission, weighted voting was adopted, which directly used weights tuned on the fold-$1$ evaluation set. However, this resulted in minimal performance improvements and was thus not considered here.

\section{Results and Discussion}
\label{sec:results}

In order to evaluate the effectiveness of the FCNN and DCNN architectures, the short-term FBank and long-term scalogram features were tested separately on both neural networks. As demonstrated in the first $4$ rows of Table \ref{tab:feat_compare}, a poorer performance was observed for FBank-DCNN and scalogram-FCNN compared with FBank-FCNN and scalogram-DCNN. This can be attributed to the convolution methods used for each architecture. FCNNs adopt $2D$ convolutional kernels that are applied to features with a strong time-scale correlation, while DCNNs adopt $1D$ convolutional kernels that are applied to features that only exhibit a scale correlation. Moreover, the ave-diff features exhibited higher accuracies compared to the left-right features. This indicates that the averaged and differential sound contains a larger amount of straight-forward scene information. For stereo audios, the sound of a passing car, for example, contains clear spatial information and the differential signal can express this more directly.

\begin{table}[tb]
	\renewcommand{\arraystretch}{1.3}
	\caption{Comparison of various short-term and long-term features on different neural network architectures.}
	\label{tab:feat_compare}
	\centering
	\resizebox{0.48\textwidth}{!}{
		\begin{tabular}{c|c|c}
			\hline
			\multirow{2}{*}{\textbf{Feature type}} & \multirow{2}{*}{\textbf{Classifier}} & \textbf{Acc.(\%)} \\
			& & \textbf{(left-right/ave-diff feature)} \\
			\hline
			Short-term FBank & FCNN & $76.92/79.95$ \\
			\hline
			Short-term FBank & DCNN & $71.42/72.00$ \\
			\hline
			Long-term Scalogram & FCNN & $76.22/80.07$ \\
			\hline
			Long-term Scalogram & DCNN & $77.54/\mathbf{82.92}$ \\
			\hline
			Long-term FBank & DCNN & $72.97/78.95$ \\
			\hline
			\multirow{2}{*}{\minitab[c]{Long-term Scalogram \\ extracted by triangle filters }} & \multirow{2}{*}{\centering{DCNN}} & \multirow{2}{*}{\centering{$\mathbf{77.68}/82.53$}} \\
			& &  \\
			\hline
			\multirow{2}{*}{\minitab[c]{Long-term Scalogram \\ w/o low-frequency filters }} & \multirow{2}{*}{\centering{DCNN}} & \multirow{2}{*}{\centering{$67.72/71.40$}} \\
			& & \\
			\hline
	\end{tabular}}
\end{table}

Besides the long-term scalogram, $3$ additional long-term features were extracted to yield the efficiency of the wavelet scales formulated in Section \ref{sec:wavelet}. A long-term FBank feature was extracted via the same window settings and the same number of filters as that of the scalogram, resulting in the same time-scale resolution. Furthermore, an additional extraction method employed triangle filters on the scale of Equation (\ref{eq:summ_lambda}). The start and end frequencies of each triangle filter were located at the center of the previous and the next triangles, respectively. The start of the first filter and the end of the last filter were set to $0$ and $f_h$, respectively. The third long-term scalogram only contained the high frequency part defined as $j=P,...,J-1$ in Equation (\ref{eq:summ_lambda}). The corresponding results are reported in the last $3$ rows of Table \ref{tab:feat_compare}. The performance of the long-term FBank did not reach that of the long-term scalogram and short-term FBank. However, with the triangle filters defined on the wavelet scale, the DCNN classifier attained similar accuracies. This indicates that the wavelet scale describes the scene information more accurately than the classic Mel scale on long-term windows, while the filter shape plays a less crucial role. In addition, after the removal of the low-frequency part, the performance degradation highlights the importance of low-frequency filters, which covered $0$ to $205Hz$ in our experiments.

\begin{figure}[tb]
	\centering
	\includegraphics[width=\linewidth]{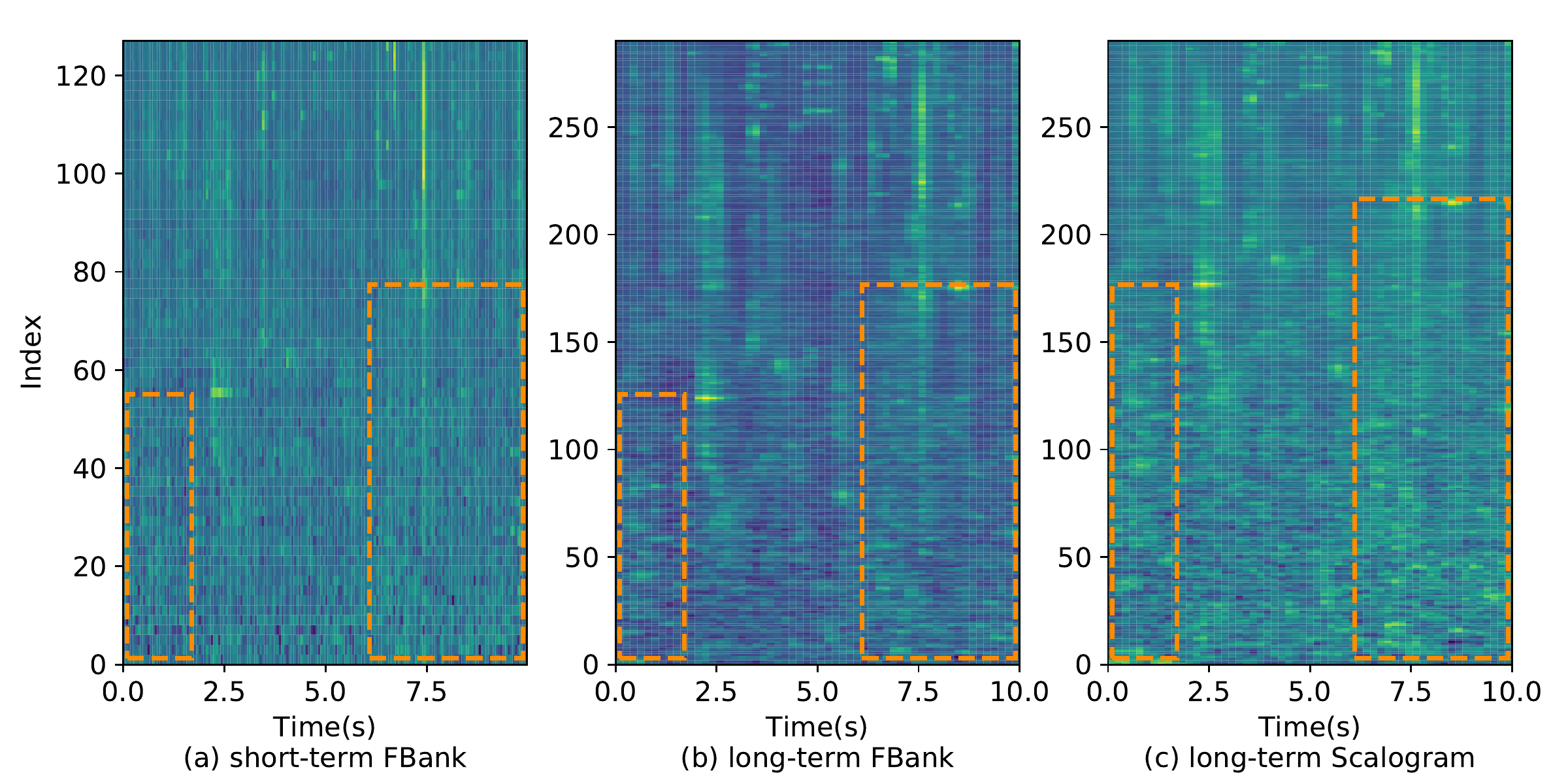}
	\caption{The FBank and scalogram extracted from the differential channel of a sample recorded in an airport. The orange box demonstrates that the feature extracted by the wavelet scale shows more high-energy details on the low frequency part.}
	\label{fig:diff_feats}
\end{figure}

In summary, the long-term ave-diff scalogram integrated with the DCNN was able to achieve the optimal performance among the different combinations of FBank, scalogram and FCNN, DCNN, exceeding the short-term left-right FBank-FCNN performance by approximately $6\%$. In order to demonstrate the differences among the short-term FBank, the long-term FBank and the long-term scalogram, we take the audio spectrum depicted in Figure \ref{fig:diff_feats} as an example. The scalogram has a lower time resolution but a higher scale resolution. Furthermore, the wavelet scale places more attention on lower frequencies, which may lead to the discovery of additional acoustic cues.

In practical terms, compared with FBank-FCNN, the scalogram-DCNN is the preferred choice for $3$ reasons. First, the scalogram occupies less storage space, with dimensions of $58\times290$, approximately $1/4$ times of $500\times128$ FBank for a $10$-second mono audio. Second, the DCNN architecture with $1D$ convolution only depends on the current frame, leading to a faster computational speed under similar numbers of network parameters compared with FCNN. Third, for real-life applications, the frame-wise predictions of acoustic scenes are more flexible than the segment-wise ones.

\begin{table*}[bt]
	\renewcommand{\arraystretch}{1.3}
	\caption{An ablation study of data augmentation with ave-diff-scalogram-DCNN classifiers.}
	\label{tab:gan_compare}
	\centering
	\resizebox{0.98\linewidth}{!}{
		\begin{tabular}{c|l|c|c|c}
			\hline
			\textbf{Augmentation method} & \qquad\qquad\qquad\quad\textbf{Settings} & \textbf{Acc.(\%) (total)} & \textbf{Acc.(\%) (unseen cities)} & \textbf{Acc.(\%) (seen cities)} \\ \hline
			- & \qquad\qquad\qquad\qquad- & $82.92$ & $67.07$ & $84.64$ \\ \hline
			\multirow{2}{*}{Speed \& volume perturb.\cite{ko2015audio}} & $3$-fold, speed factor=$0.9,1.0,1.1$,  & \multirow{2}{*}{$\mathbf{82.96}$} & \multirow{2}{*}{$64.15$} & \multirow{2}{*}{$\mathbf{85.01}$} \\ 
			& volume factor $\in [0.125,2.0]$ & & & \\ \hline
			Mixup\cite{mixup2017zhang} & $\alpha=0.1$ & $82.53$ & $68.54$  & $84.05$ \\ \hline
			SpecAug\cite{Park_2019} & $W=F=30, m_T=m_F=2, T=5$ & $81.29$ & $\mathbf{68.78}$ & $82.65$ \\ \hline
			Speed \& volume perturb. + SpecAug & the same as above & $82.94$ & $64.15$ & $84.98$ \\
			\hline\hline
			CGAN & w/o sample filter, city-based split & $82.58$ & $68.05$ & $84.16$ \\ \hline
			CGAN & w/\quad\!sample filter, city-based split & $83.03$ & $71.22$ & $84.32$ \\ \hline
			ACGAN & w/o sample filter, city-based split & $83.13$ & $70.49$ & $84.21$ \\ \hline
			ACGAN & w/\quad\!sample filter, city-based split & $\mathbf{84.06}$ & $\mathbf{73.17}$ & $\mathbf{85.25}$ \\ \hline
			ACGAN & w/\quad\!sample filter, fixed split & $82.58$ & $70.98$ & $83.84$ \\ \hline
			ACGAN & w/\quad\!sample filter, random split & $83.03$ & $71.71$ & $84.26$ \\ \hline
	\end{tabular}}
\end{table*}


\begin{table}[tb]
	\renewcommand{\arraystretch}{1.3}
	\caption{Experiments on the adversarial city training and DCT-based temporal module. The results in total, of unseen cities, of seen cities are listed with the slash.}
	\label{tab:adv_dct}
	\centering
	\resizebox{0.47\textwidth}{!}{
		\begin{tabular}{c|c|c|c}
			\hline
			\multirow{2}{*}{\textbf{Classifier}} & \multirow{2}{*}{\textbf{Data aug.}} & \textbf{Acc.(\%)} & \textbf{Acc.(\%)} \\ 
			& & \textbf{ave-diff scalogram} & \textbf{left-right Scalogram} \\ \hline 
			DCNN & \texttimes & $82.92/67.07/84.64$ & $77.54/65.61/78.83$ \\ \hline
			DCNN & \checkmark & $84.06/73.17/85.25$ & $78.81/65.61/80.24$ \\ \hline
			DCNN & \multirow{2}{*}{\checkmark} & \multirow{2}{*}{$83.58/74.63/84.56$} & \multirow{2}{*}{$79.00/65.37/80.48$} \\
			+DCT & & & \\ \hline
			DCNN & \multirow{2}{*}{\texttimes} & \multirow{2}{*}{$80.76/68.78/82.07$} & \multirow{2}{*}{$76.54/61.95/78.12$} \\
			+adv & & & \\ \hline
			DCNN & \multirow{2}{*}{\checkmark} & \multirow{2}{*}{$84.16/70.98/\mathbf{85.59}$} & \multirow{2}{*}{$\mathbf{79.90}/72.68/\mathbf{80.69}$} \\
			+adv & & & \\ \hline
			DCNN & \multirow{3}{*}{\checkmark} & \multirow{3}{*}{$\mathbf{84.23}/\mathbf{75.61}/85.17$} & \multirow{3}{*}{$79.86/\mathbf{75.12}/80.37$} \\ 
			+adv & & & \\
			+DCT & & & \\ \hline
	\end{tabular}}
\end{table}

The ave-diff-scalogram-DCNN classifier was chosen to evaluate the data augmentation scheme. We first conducted experiments on the classic data augmentation techniques (row 1-5 in Table \ref{tab:gan_compare}). The speed and volume perturbation \cite{ko2015audio} generated duplicated samples with different speed and volume factors. It improved the accuracy on the seen cities by training on more source data but failed on unseen cities. The Mixup \cite{mixup2017zhang} and SpecAug \cite{Park_2019} interpolated and masked input features to improve the system's generalization on unseen cities, but the accuracies decreased on seen cities. We tried to combine speed and volume perturbation with SpecAug, but no further improvement occurred.

\begin{figure}[b]
	\centering
	\includegraphics[width=\linewidth]{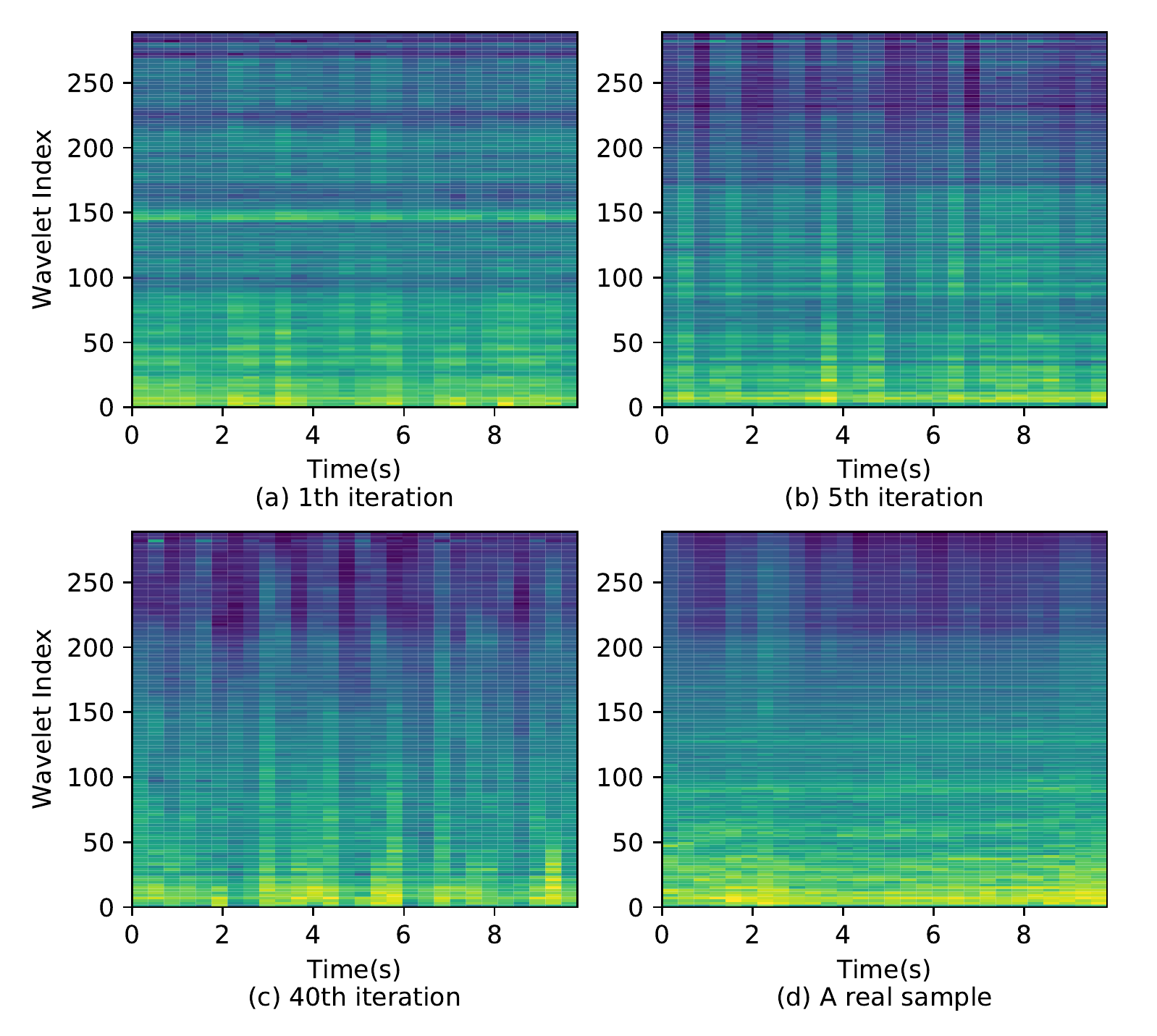}
	\caption{The evolution of the generated differential-channel scalograms. The frame-wisely generated samples do not exhibit temporal correlation, but this shortcoming does not affect the DCNN classifier.}
	\label{fig:gan_evolution}
\end{figure}

At each iteration of the proposed GAN-based data augmentation, dataset split, the ACGAN and the neural network-based sample filter play crucial roles in generating fake feature samples. An ablation study of these modules was performed, with results reported in Table \ref{tab:gan_compare}. With the absence of the sample filter, the fake features generated by CGAN were not able to improve the performance of the final classifier, despite the accuracy improvements in the testing subset. By replacing CGANs with ACGANs and adding the sample filter, the final classifier was able to perform better. The ACGAN-based data augmentation with the sample filter exhibited the highest accuracies both on the records from seen and unseen cities. In particular, a relative error rate reduction of $6.7\%$ compared with the original ave-diff-scalogram-DCNN system was observed. The evolution of the generated fake scalograms is depicted in Figure \ref{fig:gan_evolution}. The quality of the generated data increased following the iterations, yet it was difficult to determine whether it benefited the final classifier, thus requiring the application of the sample filter. 


Three methods were employed to split the datasets during ACGAN training, as listed in the last $3$ rows in Table \ref{tab:gan_compare}. Results demonstrate that the splitting method was crucial to final performance. Although the fixed and random split methods improved the accuracies for the unseen city records, they failed to classify audios from the seen cities. The reason might be that the city-based dataset split optimized the classifiers more robust on the seen city while the data augmentation improves the performance for the unseen cities records. In summary, $3$ most important modules of data augmentation are addressed, which are the city-based dataset split, the ACGAN architectures and the sample filter.

The adversary training branch was added to the $clf^A, clf^B$ and final classifier. As shown in Table \ref{tab:adv_dct}, its performance was inferior to the original classifier (row $1$ and $4$). Improvements were observed for the extended dataset (row $2-3$ and $5-6$). The results can be attributed to the potential of the adversary training to improve the system performance, yet it should be adopted under data augmentation.

Moreover, the DCT-based temporal module was tested on the final classifiers (Table \ref{tab:adv_dct}). The temporal module was observed to dramatically decrease the loss on the training and validation, however overfitting might be introduced. Despite inconsistencies in improvements, this approach was adopted for the diversity of the fusion systems.

The highest accuracy without the system fusion was $84.23\%$ on the fold-$1$ evaluation set with the frame-wise ACGAN-based data augmentation scheme, the ave-diff scalogram as the input feature, the DCNNs as the classifiers, combined with the adversarial city training and the DCT-based temporal module.

The data augmentation scheme was also tested under the short-term FBank scheme. The generator and discriminator for FBank adopted the architectures described in Tables \ref{tab:seg_gen}. The proposed data augmentation scheme was observed to slightly improve accuracies (Table \ref{tab:feat_compare} and \ref{tab:fusion_compose}). Considering that our baseline was able to achieve high accuracies, the performance improvement of the proposed data augmentation scheme was stable and considerable.

The short-term FBank and long-term scalogram compliments each other. A fusion ratio of $1:2$ yielded optimal results in our experiments. Table \ref{tab:fusion_compose} reports the fusion results of our submission for the DCASE2019 challenge on the fold-$1$ evaluation set.

In order to compare our proposed scheme with the SVM-GAN-based scheme in \cite{Mun2017}, the fusion systems were trained and tested using the DCASE17 dataset. The majority of the DCASE19 settings remained unchanged, with the exception of the slight modification of the DCNNs and FCNNs for this bit small dataset. The number of hidden units $h$ for DCNN was $512$ (Table \ref{tab:network_params}), while Conv7 and Conv8 (Table \ref{tab:fcnn}) were removed. 
As no city information was revealed, the dataset was split according to the recording index and adversarial training was not performed. The detailed results are reported in Table \ref{tab:fusion_compose}. The scalogram-DCNN architecture combined with the proposed data augmentation scheme achieved an accuracy of $82.72\%$. After fusion, the accuracy increased to  $84.57\%$, exceeding the top systems in the competition by $1.3\%$ \cite{Mun2017}.

\begin{figure}[h]
	\centering
	\includegraphics[width=\linewidth]{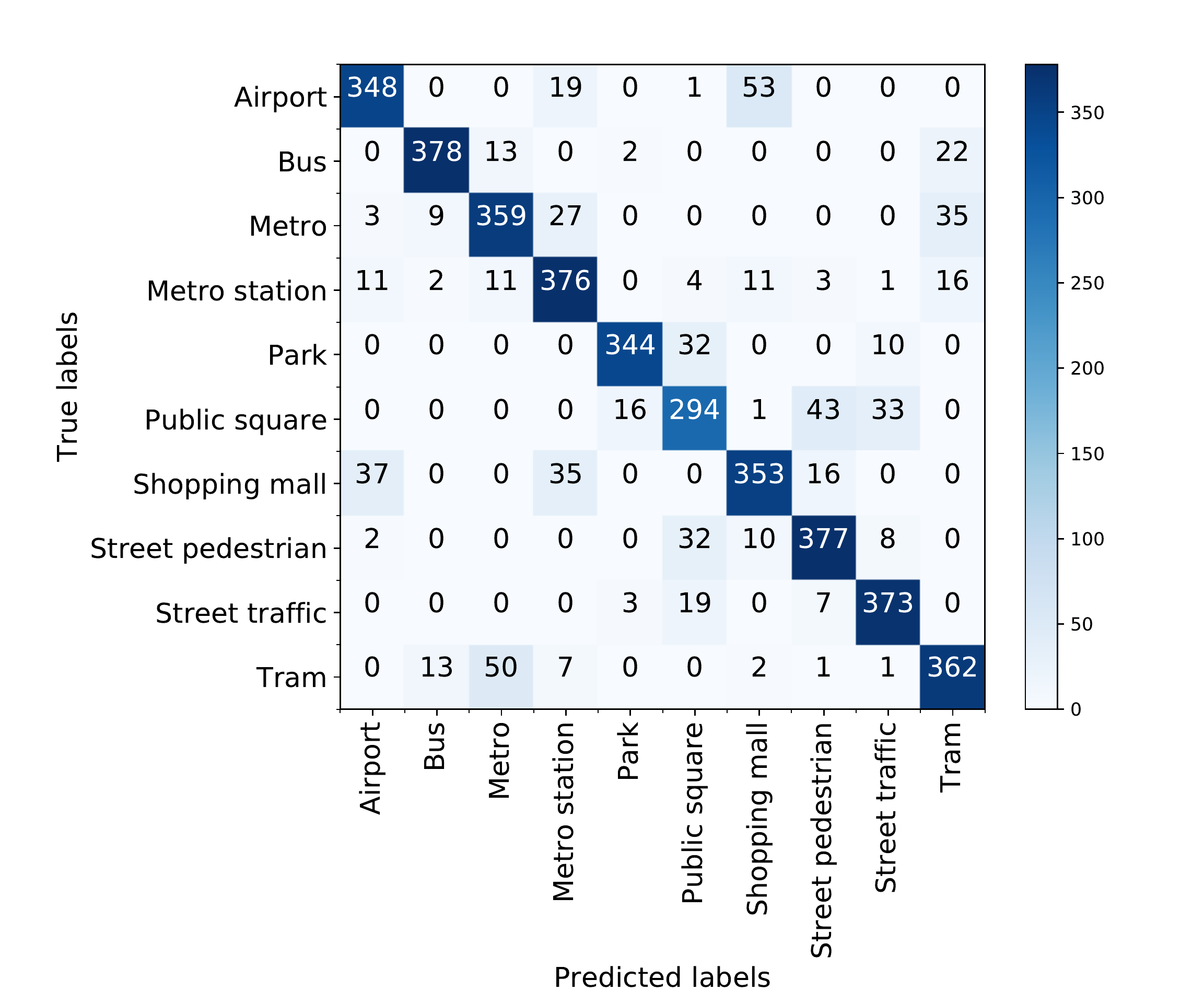}
	\caption{The confusion matrix of the fusion system for DCASE19 Task 1A.}
	\label{fig:confusion}
\end{figure}

The confusion matrix of the DCASE19 fusion system is presented in Figure \ref{fig:confusion}. Some scenes that are confused by classifiers are also difficult for humans to process, such as public squares and street pedestrians, or airports and shopping malls.

\begin{table}[tb]
	\renewcommand{\arraystretch}{1.3}
	\caption{System fusion for DCASE19 and DCASE17 dataset, where ACGAN-based data augmentation was adopted for all classifiers.}
	\label{tab:fusion_compose}
	\centering

	\begin{tabular}{c|c|c|c|c}
		\hline
		\multirow{2}{*}{\textbf{Feature type}} & \multirow{2}{*}{\textbf{Channel}} & \multirow{2}{*}{\textbf{Classifier}} & \textbf{DCASE19} & \textbf{DCASE17} \\ 
		& & & \textbf{Acc.(\%)} & \textbf{Acc.(\%)} \\ \hline
		Long-term & \multirow{2}{*}{ave-diff} & \multirow{2}{*}{DCNN} & \multirow{2}{*}{$84.06$} & \multirow{2}{*}{$\mathbf{82.72}$} \\ 
		 scalogram & & & & \\ \hline
		Long-term & \multirow{2}{*}{ave-diff} & DCNN & \multirow{2}{*}{$83.58$} & \multirow{2}{*}{$81.73$} \\
		 scalogram & & +DCT & & \\ \hline
		Long-term & \multirow{2}{*}{ave-diff} & DCNN & \multirow{2}{*}{$84.16$} & \multirow{2}{*}{-} \\
		 scalogram & & +adv & & \\ \hline
		Long-term & \multirow{2}{*}{ave-diff} & DCNN & \multirow{2}{*}{$\mathbf{84.23}$} & \multirow{2}{*}{-} \\ 
		 scalogram & & +adv+DCT & & \\ \hline
		Short-term  & \multirow{2}{*}{left-right} & \multirow{2}{*}{FCNN} & \multirow{2}{*}{$80.10$}  & \multirow{2}{*}{$71.11$} \\ 
		FBank & & & & \\ \hline
		Short-term & \multirow{2}{*}{ave-diff} & \multirow{2}{*}{FCNN} & \multirow{2}{*}{$77.56$}  & \multirow{2}{*}{$72.53$} \\
		FBank & & & & \\ \hline
		\multicolumn{3}{c|}{Average voting} & $\mathbf{85.16}$ & $\mathbf{84.57}$ \\ \hline
	\end{tabular}
\end{table}

\section{Conclusion}
\label{sec:conclusion}

We propose a novel ACGAN-based data augmentation scheme for acoustic scene classification. Integrated with the long-term scalogram extracted by the designed wavelet scale, this framework can achieve high accuracies on the testing samples, even for those recorded in unseen cities. However, the filters are still hand-tailored, with more advanced scales existing for ASC tasks. In the future, we will focus on automatically training filters without the occurrence of overfitting. 



%


\ifCLASSOPTIONcaptionsoff
  \newpage
\fi


\bibliographystyle{IEEEtran}
\bibliography{refs}




\end{document}